\documentclass{amsart}

\usepackage{graphicx}

\usepackage{color}

\usepackage{amsmath}

\makeatletter
\newcommand{\addresseshere}{%
  \enddoc@text\let\enddoc@text\relax
}
\makeatother

\newcommand{\bbm}{\begin{bmatrix}}
\newcommand{\bpm}{\begin{pmatrix}}
\newcommand{\ebm}{\end{bmatrix}}
\newcommand{\epm}{\end{pmatrix}}

\newcommand{\ds}{\displaystyle}

\newcommand{\ddx}[2]{\frac{d #1}{d #2}}
\newcommand{\ddt}[1]{\frac{d #1}{dt}}

\newcommand{\dsddt}[1]{\displaystyle\frac{d #1}{dt}}

 \newcommand{\dsdel}[2]{\displaystyle\frac{\partial #1}{\partial #2}}

\usepackage[margin=1in,footskip=0.25in]{geometry}

\begin{document}

\title[Infectious Recovery and Online Social Networks]{Analysis of Infectious-Recovery Epidemic Models for Membership Dynamics of Online Social Networks}

\author[Daniel Cooney]{Daniel Cooney\textsuperscript{1,2}}
\author[Francisco Prieto]{Francisco Prieto-Castrillo \textsuperscript{1,3}}
\author[Yaneer Bar-Yam]{Yaneer Bar-Yam \textsuperscript{1,3}}

\address{\textsuperscript{1} New England Complex Systems Institute, 110 Broadway, Suite 101,
  Cambridge, MA 02139, USA.}

\address{\textsuperscript{2} Program in Applied and Computational Mathematics, 
  Princeton University. Princeton, NJ, 08544, USA.}

\email{dcooney@math.princeton.edu}

\address{\textsuperscript{3} Media Lab, Massachusetts Institute of Technology, Cambridge, MA, 02139, USA. }


\begin{abstract}
The recent rapid growth of social media and online social networks (OSNs) has raised interesting questions about the spread of ideas and fads within our society. In the past year, several papers have drawn analogies between the rise and fall in popularity of OSNs and mathematical models used to study infectious disease. One such model, the irSIR model, made use of the idea of "infectious recovery" to outperform the traditional SIR model in replicating the rise and fall of MySpace and to predict a rapid drop in the popularity of Facebook. Here we explore the irSIR model and two of its logical extensions and we mathematically characterize the initial and long-run behavior of these dynamical systems. In particular, while the original irSIR model always predicts extinction of a social epidemic, we construct an extension of the model that matches the exponential growth phase of the irSIR model while allowing for the possibility of an arbitrary proportion of infections in the long run.
\end{abstract}
\maketitle   
\section{Introduction}
The rising popularity of online social networks (OSNs) raises many interesting questions about the nature of human social dynamics. OSNs provide a new, expedient medium for the transfer of ideas and have facilitated the emergence of new social phenomena like memes \cite{shifmanmemes} and viral videos \cite{cintron2014go}. In addition, the dynamics of growth and decline of OSN membership themselves provide an interesting setting to explore human behavior and the social spread of ideas. In this paper, we explore the persistence of OSN membership using models that draw their inspiration from epidemiology, the study of the spread of disease within a population. Recent works based upon such models have suggested that membership will decrease and disappear not long after the initial growth period \cite{cannarella2014epidemiological,zhu2014demographic}. The long time persistence of fads is an important aspect of understanding their dynamical behavior and the practical aspects of investment and planning of corporations that build and rely upon their presence.

Epidemic models have long been used to study the dynamics of the internet and the spread of ideas more broadly. In the 1960s, Daley and Kendall used a variant of the Susceptible-Infectious-Removed (SIR) model to study the spread of rumors \cite{daley1964epidemics}, and in the past decade Vespignani and Pastor-Satorras examined epidemic thresholds to study the vulnerability of systemic failure of the internet \cite{vesppas} and Jeffs et al used epidemic models to study the growth and decline of political parties \cite{jeffs2016activist}. Bettencourt et al applied ODE models of epidemics to study the adoption of new ideas \cite{bettencourt2006power} and Jin et al used similar models to study the spread of news on Twitter \cite{jin2013epidemiological}. In early 2014, Cannarella and Spechler \cite{cannarella2014epidemiological} and Ribiero \cite{ribeiro2014modeling} separately developed ODE models for dynamics of OSN membership with inspiration from mathematical epidemiology. Around the same time, Cintron-Arias used a similar model to study the spread of viral videos \cite{cintron2014go} and Bauckhage et al used epidemic models to study the spread of memes \cite{bauckhage2011insights}.

The Cannarella and Spechler model was based on the SIR model, a canonical epidemic model first proposed by Kermack and McKendrick in 1927 \cite{kermack1927contribution}. In the SIR model, each member of a population of constant size $N$ exists in one of three states: Susceptible (S), Infectious (I), or Recovered (R). In the traditional SIR model, susceptible individuals become infectious through interactions with infectious individuals, and infectious individuals progress to a recovered state. The population is assumed to be well-mixed, so interactions between susceptible and infectious individuals (and the corresponding event of infection) take place with probability proportional to the product of the fractions of individuals in the two states. For more information on modeling the spontaneous and infectious reactions in epidemic processes, see Perra et al \cite{perra2011towards}. 

\begin{figure}[h!]
\includegraphics[scale=0.3]{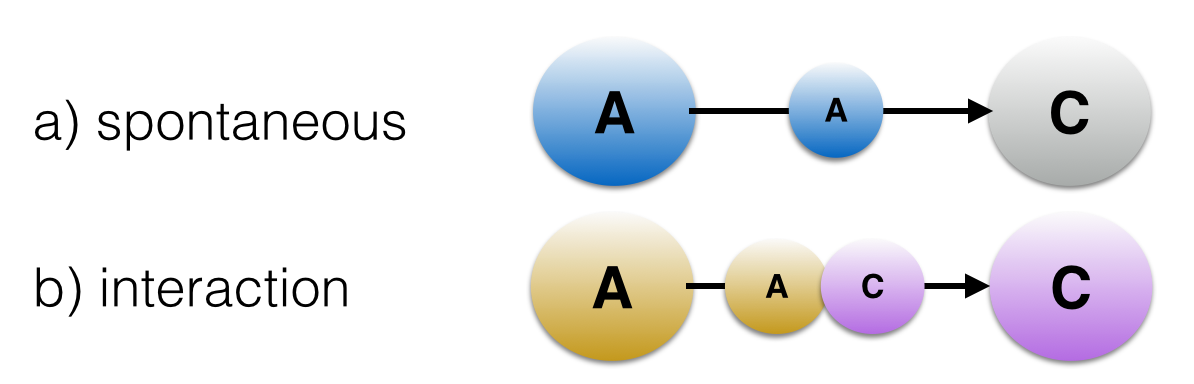}
\caption{Legend for Schematic Diagrams: Two basic elements we use to construct schematic representations of epidemic models. a) spontaneous transition from $A$ to $C$. b) infectious transition from $A$ to $C$, where type $A$ individual makes the transition due to an interaction with a type $C$ individual. We use concatenations of these motifs to represent the possible spontaneous and infectious transitions for a given epidemic process.}
\label{fig:legend}
\end{figure}

\begin{figure}[h!]
\includegraphics[scale=0.5]{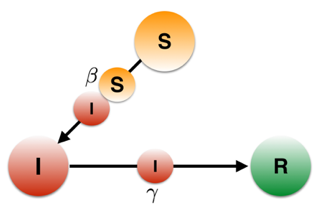}
\caption{Schematic Representation of the SIR Model:  The $S \to I$ transition occurs through an interaction between an $S$ and an $I$, whereas the $I \to R$ transition is spontaneous.}
\label{fig:SIRschematic}
\end{figure}

One can represent the states and state transitions of the SIR model with the schematic diagrams in Figs. \ref{fig:legend} and \ref{fig:SIRschematic}, where arrows going through pairs of nodes represent reactions that result from interactions, and arrows going through a single node corresponds to a spontaneous transition. We can represent the changing number of S, I, and R individuals through the nonlinear system Eq. \ref{eq:SIR}. Note that population size $N$ is constant. Because this will be true of each of the models we study in the paper, we can divide both sides of our differential equations by $N$ to obtain equations for the changing fractions $\frac{S}{N}$, $\frac{I}{N}$, and $\frac{R}{N}$ within the population. From Section 2 onward, we will consider differential equations in terms of these proportions, but for convenience we will maintain the notation of $S$, $I$, and $R$ to represent these fractions.

\begin{equation} \label{eq:SIR}
\begin{array}{ll}
\dsddt{S} &= - \ds\frac{\beta SI}{N}  \vspace{2mm}  \\ 
\dsddt{I} &= \ds\frac{\beta SI}{N}  - \gamma I   \vspace{2mm} \\ 
\dsddt{R} &= \gamma I 
\end{array}
\end{equation}

\vspace{5mm}

\begin{figure}[h!]
\includegraphics[scale=0.5]{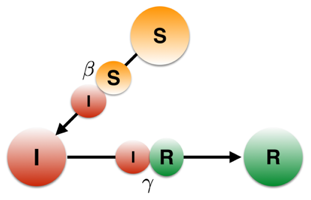}
\caption{Schematic Representation of the irSIR Model: The $S\to I$ and $I \to R$ transitions are both infectious.}
\label{fig:irSIRschematic}
\end{figure}

Cannarella and Spechler propose a modification to the traditional SIR model known as ``infectious-recovery," in which an individual's recovery transition from $I$ to $R$ does not occur spontaneously, but rather due to an interaction with a recovered individual \cite{cannarella2014epidemiological}. In this sense, recovery is itself an infectious process, as recovery is spread through the social interaction of infectious and recovered individuals. The intuition behind this modification is that individuals join OSNs because their real-life friends are also members (the S-I interaction) and then individuals remain members until they see their friends have left (the I-R interaction). In the context of OSN membership, Cannarella and Spechler interpret the susceptible ($S$) class to be individuals who have yet to join the OSN, infectious individuals ($I$) are active users of the OSN, and recovered ($R$) individuals are former users of the OSN. Their infectious-recover SIR (irSIR) model can be represented by the schematic in Fig. \ref{fig:irSIRschematic} or through the system of ODEs in Eq. \ref{eq:irSIRpopN}.

\begin{equation} \label{eq:irSIRpopN}
\begin{array}{ll}
\dsddt{S} &= - \ds\frac{\beta SI}{N}  \vspace{2mm}  \\ 
\dsddt{I} &= \ds\frac{\beta SI}{N}  - \ds\frac{\gamma IR}{N}   \vspace{2mm} \\ 
\dsddt{R} &= \ds\frac{\gamma IR}{N} 
\end{array}
\end{equation}

Cannarella and Spechler apply their model to study data from Google Trends for the queries ``MySpace" and ``Facebook" \cite{cannarella2014epidemiological}. Using best-fit techniques, they found parameters for the irSIR model which captured the initial exponential growth and subsequent exponential decay of Google Search interest in MySpace, showing that the irSIR model was more succesful than the SIR model in reproducing the empirical behavior. They apply the same best-fit techniques to search data for Facebook, which has yet to experience a decline similar to MySpace. The paper extrapolated from the best-fit curves to predict that Facebook could lose up to 80 percent of its users in the near future \cite{cannarella2014epidemiological}.

Several papers have extended the work of Cannarella and Spechler. Fuji uses an asymptotic expansion to find solutions for $S(t)$, $I(t)$, and $R(t)$ in the exponential growth phase of the irSIR model \cite{fujii2014comment}. Instead of studying the dynamics of a single OSN in isolation, Tanaka et al consider the takeover of an incumbent social network (users in state $I$), by a new social network (users in state $R$). 
They use Google Trends data 
to demonstrate that the irSIR model provides a better fit than the SIR model in describing the rise of Facebook's popularity in Japan at the expense of the incumbent OSN Mixi \cite{tanaka2015study}. 
Zhu et al study an ``irSIRS'' model, extending the irSIR model to include demographic terms (corresponding to the creation and deletion of "bot" accounts) as well as the possibility for recovered individuals to once again become susceptible. They fit their model to user data for the Chinese social networks Renren and Sina-Weibo and predict that the two OSNs will also experience a near-term downturn in active membership \cite{zhu2014demographic}.

Cintron-Arias uses an epidemic model with an infectious-recovery term to describe the popularity of viral videos. He shows that the model considered exhibits exponential growth for any positive contact rate $\beta$ and validates the model with data of view counts of popular Youtube videos \cite{cintron2014go}. 
Ribeiro conducts stability analysis for an OSN model in which active users become inactive due to spontaneous reactions and inactive users can return to active status through interaction with active users  \cite{ribeiro2014modeling}.
In this model, Ribeiro discovers a phase transition between a regime in which all users ultimately become inactive and one in which a nonzero fraction of users remain active in the long-run.

Here we consider the question of long term behavior for infectious-recovery epidemic models.
We show that the long-run behavior of the irSIR model will always result in full abandonment of an OSN. This leads to an interesting question: are there logical extensions of the irSIR model which can result in an asymptotically stable endemic equilbrium? 
We take inspiration from several biological mechanisms through which a long-run endemic equilibrium can be sustained in extensions of the traditional SIR model: waning immunity and demographic turnover \cite{keeling2008modeling}. 
We show that these reasonable augmentations to the dynamical process of infectious-recovery epidemics lead to dramatic changes in the long time behavior. Characterizing these processes should be helpful for understanding the dynamics of social phenomena and for providing numerical tools to analyze underlying patterns in abundantly available social media data.

The paper is structured as follows. In Section 2, we will analyze Cannarella and Spechler's original irSIR model through both stability analysis of the system's equilibria and by evaluating the peak infection size. Then, in Sections 3 and 4, we analyze two extensions of the irSIR model. These extensions are respectively named the irSIRS model with spontaneous loss of immunity (recovered individuals can return to susceptibility on their own), 
and the demographic irSIR model (in which new susceptibles are introduced to the population at the same rate as individuals of each type are removed).

\section{Analysis of \lowercase{ir}SIR Model}

We consider the system 
\begin{equation} \label{eq:irSIR}
\begin{array}{ll}
\dsddt{S} &= - \beta SI \vspace{2mm}  \\ 
\dsddt{I} &= \beta SI  - \gamma IR   \vspace{2mm} \\ 
\dsddt{R} &= \gamma IR 
\end{array}
\end{equation}
with $S+I+R = 1$, $S,I,R \geq 0$, and $\beta, \gamma > 0$. 
We identify points on the simplex of allowed values using the notation $(S,I,R)$.

\subsection{Equilibrium Analysis}

We begin by reviewing the steady states of the dynamical system. By inspection, all uninfected states (with $I=0$) are in equilibrium. Therefore any point with the form $(x, 0, 1-x)$ for $x \in [0,1]$ is a fixed point. From the first equation either $S$ or $I$ must equal 0 if $\ddt{S} = 0$, and either $I$ or $R$ must equal 0 from the third equation. It follows that the point $(0,1,0)$ is the only additional equilibrium.
We can represent the three-dimensional system by the two equations

\begin{equation} \label{eq:irSIRreduced}
\begin{array}{ll}
\dsddt{S} &= - \beta SI \vspace{2mm}  \\ 
\dsddt{I} &= (\beta + \gamma) SI  - \gamma( I - I^2)  \vspace{2mm}
\end{array}
\end{equation}
by inserting $R = 1 - S - I$. The system has Jacobian matrix $$J(S,I)  = \begin{pmatrix} -\beta I & -\beta S \\ (\beta + \gamma) I &  \: \: \: \: (\beta + \gamma) S - \gamma(1 - 2I) \end{pmatrix}$$

At the all-infectious equilibrium $(0,1,0)$, the Jacobian 
has eigenvalues $\gamma$ and $-\beta$, so this equilibrium 
is unstable. $J(x,0)$ has eigenvalues of 
$0$ and $ \beta x + (x - 1) \gamma$, the latter of which is positive when
$x > \frac{\gamma}{\beta + \gamma}$.
Thus equilibria of the form $(x,0,1-x)$ are unstable for $x > \frac{\gamma}{\beta + \gamma}$, so a small injection of infectious individuals introduced into such a population will result in an initial exponential growth of cases.

\subsection{Maximum Number of Infectious Individuals}

An interesting difference between the irSIR and SIR models is the range of possible values for $I(t)$ when the dynamics start from a small perturbation of the disease-free equilibrium. The maximum number of infectious individuals can reach an arbitrarily large proportion of the populations in the irSIR model. This makes the model particularly relevant for applications to internet tools and memes, as, for example, the complete saturation of the population should be a realistic possibility for tools such as email. 
 We show this using Eq. \ref{eq:irSIR} to obtain

\begin{figure}[h!]
\includegraphics[angle=0,scale=0.45]{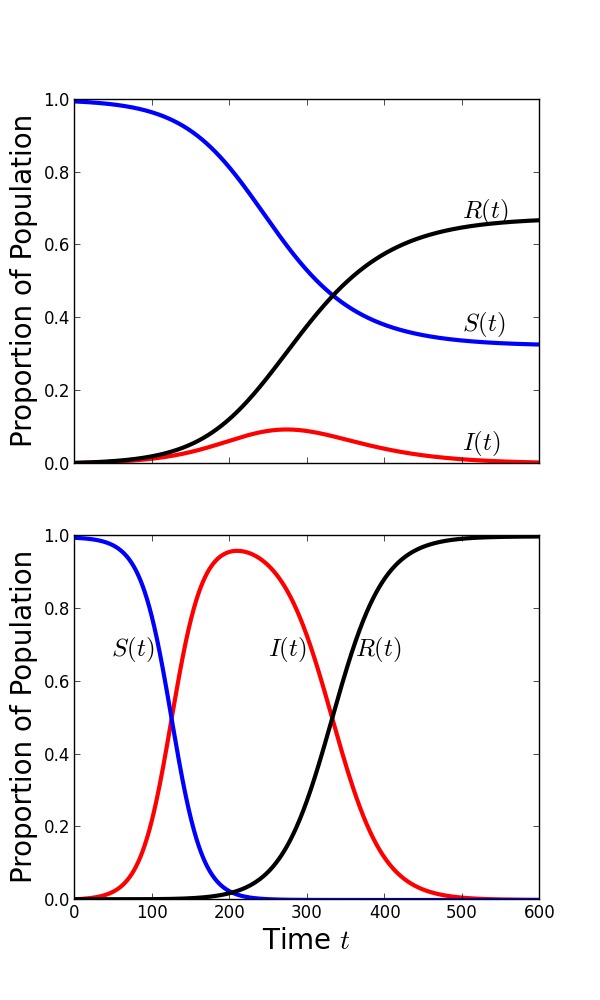}
\label{fig:sirmodelmax}
\caption{Numerical integration of SIR and irSIR models with curves representing $S(t)$ (blue), $I(t)$ (red), and $R(t)$ (black) individuals in the population. These dynamics were run with parameter values $\beta = 0.05$ and $\gamma = 0.03$, and initial proportions of $S(0) = 0.996$ and $I(0) = R(0) = 0.002$. Observe the different peaks of $I(t)$ for the two models.}
\end{figure}

\begin{figure}[h!]
\includegraphics[angle=0,scale=0.45]{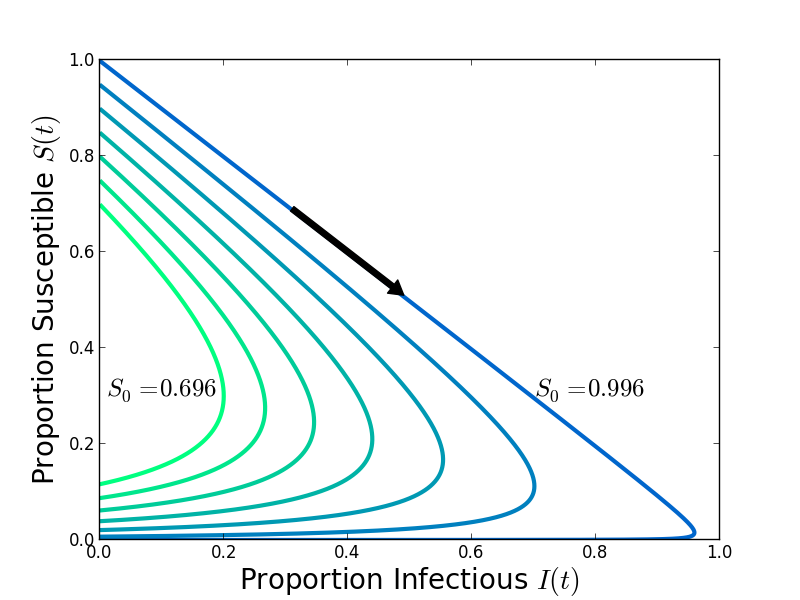}
\label{fig:irSIRphasespace}
\caption{Trajectories of irSIR System on the simplex of possible states: trajectories for the irSIR system with initial condition near a point $(x,0,1-x)$ for $x > \frac{\gamma}{\beta + \gamma}$. The curves display the initial exponential growth in $I(t)$ and subsequent exponential decline paired with continual decrease in the number of susceptibles $S(t)$. This corresponds to initial rightward motion and subsequent leftward motion in the $I(t)$ direction and downward motion for all time in the $S(t)$ direction. The black arrow indicates the counter-clockwise evolution of the system over time.}
\end{figure}
$$\ddx{R}{S}
= - \frac{\gamma}{\beta} \frac{R}{S}$$
so
 \begin{equation} \label{eq:SRrelation} R(t) = R(0) \cdot S(0)^{\displaystyle\frac{\gamma}{\beta}}\cdot S(t)^{- \displaystyle\frac{\gamma}{\beta}} \end{equation}

If $S(0) < \frac{\gamma}{\beta + \gamma}$, a small $I(0)$ results in an initial exponential growth in $I(t)$, which is maximized ( $\ddt{I} = 0$)
when $S = \ds\frac{\gamma}{\beta} R$. From Eq. \ref{eq:SRrelation} at peak infectiousness ( time $t^*$), 
\begin{equation} S(t^*) = \left( \frac{\gamma}{\beta} \left(S(0)\right)^{\displaystyle\frac{\gamma}{\beta}} \left(R(0)\right)
\right)^{\displaystyle\frac{\beta}{\beta + \gamma}}
\: \: \mathrm{and} \: \:
 I(t^*) = 1 -  \frac{\gamma + \beta}{\gamma} \left( \frac{\gamma}{\beta} \left(S(0)\right)^{\displaystyle\frac{\gamma}{\beta}} \left(R(0)\right) 
\right)^{\displaystyle\frac{\beta}{\beta + \gamma}} \end{equation}

\begin{figure}[h!]
\includegraphics[scale=0.5]{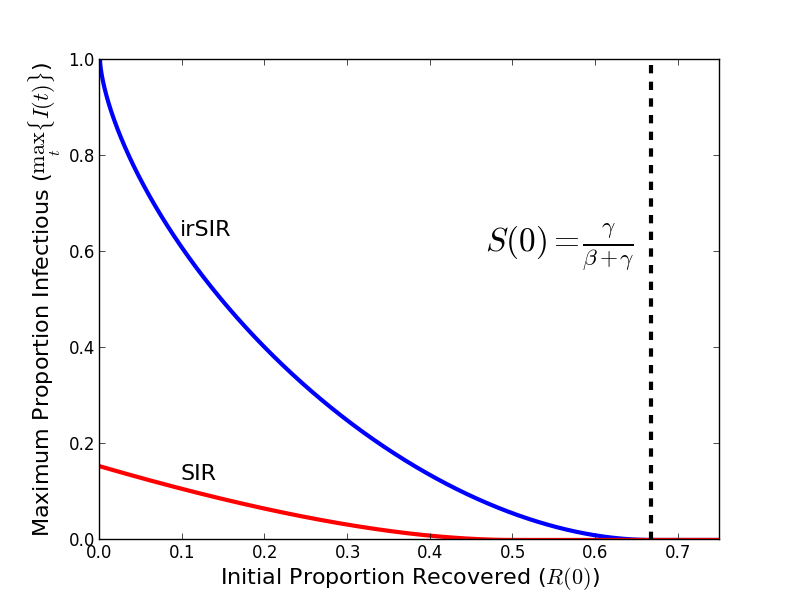}
\caption{Maximum number of infectious individuals for both the SIR and irSIR models based upon peak infectiousness formulas ($\beta = 0.5$, $\gamma = 0.25$) with varying values of $R(0)$, the initial proportion of recovered individuals, and an initially-seeded infectious class of proportion $I(0) = 0.0001$.}
\label{fig:maxinfections}
\end{figure}

On the other hand, if $S(0) < \frac{\gamma}{\beta + \gamma}$ ($R(0) > \frac{\gamma}{\beta + \gamma}$ for small enough $I(0)$), we have $$\ddt{I} = \beta S I - \gamma I R < I \left( \beta \frac{\gamma}{\beta + \gamma} - \gamma \frac{\beta}{\beta + \gamma}  \right) = 0 $$ for all time $t$ and therefore the peak infectiousness will equal $I(0)$. 
For the SIR model, one can use a similar approach to derive the maximum
$$ I_{MAX} = I(0) + S(0) - \frac{\gamma}{\beta} \left(1 - \log{\left( \frac{\gamma}{\beta} \right) + \log{\left(S(0)\right)}}\right)$$ 
when $S(0) < \frac{\gamma}{\beta}$ and $I_{MAX} = I(0)$ otherwise \cite{weiss}.
\hspace{0.5mm} It follows that the peak infectiousness is achieved at $t = 0$.
The values of $I(t^*)$ and $I_{MAX}$ are displayed in Fig. \ref{fig:maxinfections} for various initial proportions of recovered individuals $R(0)$. For this set of parameter values $\beta$, $\gamma$, and $I(0)$, we see that $I(t^*) > I_{MAX}$ and 
$I(t^*) \approx 1$ when $R(0) \to 0$ and $I(t^*) = I(0) \approx 0$ when $R(0) \to \frac{\beta}{\beta + \gamma}$.

We can also obtain the total number of individuals that become infected over the course of the outbreak \cite{weiss}. Because $I(t)$ vanishes in the long run, we obtain the relationship $S(\infty) = 1 - R(\infty)$ as $t \to \infty$, so we can numerically solve for $R(\infty)$ by substituting for $S(\infty)$ in Eq.  \ref{eq:SRrelation}. Then subtracting $R(0)$ from $R(\infty)$ gives us the total number of individuals that were ever in state $I$ at any point in the outbreak.

\newpage

\section{\lowercase{ir}SIRS Model with Spontaneous Loss of Immunity}

\begin{figure}[h!]
\includegraphics[scale=0.6]{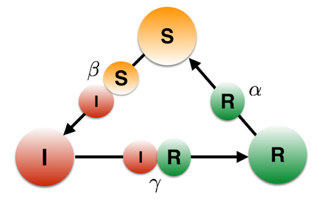}
\caption{Schematic Representation of irSIRS Model with Spontaneous Loss of Immunity: The $S\to I$ and $I \to R$ transitions are infectious, but the $R \to S$ transition is spontaneous.}
\label{fig:irSIRSsponschematic}
\end{figure}

We now consider an extension of the irSIR model in which individuals in the removed class can once again become susceptible through a spontaneous reaction (Fig. \ref{fig:irSIRSsponschematic}). This models individuals who have left the social network transitioning (without interaction) to a state in which they can be convinced to the rejoin through interaction with an active user. This is given by the set of equations

\begin{equation} \label{eq:irSIRSspon}
\begin{array}{ll}
\dsddt{S} &= - \beta SI + \alpha R \vspace{2mm}  \\ 
\dsddt{I} &= \beta SI  - \gamma IR   \vspace{2mm} \\ 
\dsddt{R} &= \gamma IR - \alpha R
\end{array}
\end{equation}

where parameters satisfy $\beta, \gamma > 0$ and $\alpha \geq 0$ ($\alpha = 0$ corresponds to the irSIR model). There are two equilibria at the corners of the simplex: the disease-free equilibrium $(1,0,0)$ and the all infectious equilibrium $(0,1,0)$. If $\alpha < \gamma$, there is an interior equilbrium  $$(S^*_{e},I^*_{e}, R^*_{e}) = \left( \displaystyle\frac{\gamma - \alpha}{ \beta + \gamma}, \displaystyle\frac{\alpha}{\gamma}, \displaystyle\frac{\beta \left(\gamma - \alpha \right)}{ \gamma \left(\beta + \gamma \right) } \right)$$ 

\subsection{Local Stability Analysis}
\begin{figure}[h!]
\includegraphics[scale=0.45]{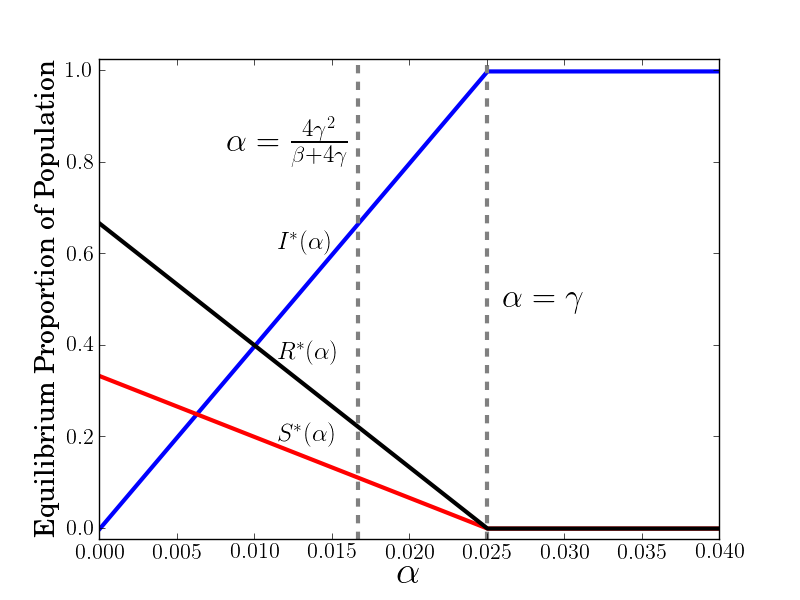}
\caption{Stable Equilibria for irSIRS Model with Spontaneous Loss of Immunity: Here we display the values of $S(t)$ (red), $I(t)$ (blue), and $R(t)$ (black) at the unique stable equilibrium for given $\alpha$. The right dotted line corresponds to the bifurcation point $\alpha = \gamma$ where we transition from a stable interior equilibrium to a stable all-infectious equilibrium. The left dotted line corresponds to the point $\alpha = \frac{4 \gamma^2}{\beta + 4 \gamma}$, which is the value of $\alpha$ at which the interior equilibrium transitions from being a stable spiral to being a stable node. }
\label{fig:irSIRSsponbifurcation}
\end{figure}

For the equilibrium analysis, we use the conservation of population to write
\begin{equation} \label{eq:irSIRSsponsiform}
\begin{array}{ll}
\dsddt{S} &= -  \left(\alpha + \beta I \right)S + \alpha \left( 1 - I \right) \vspace{2mm}  \\ 
\dsddt{I} &=  \hspace{4mm} \left(\beta + \gamma \right) SI - \gamma \left(I - I^2 \right) \vspace{2mm} \\ 
\end{array}
\end{equation}
with Jacobian $$J(S,I) = \begin{pmatrix} -\left( \alpha + \beta I \right) & - \left( \alpha + \beta S \right) \\ \left(\beta + \gamma \right) I & \left(\beta + \gamma \right)S - \gamma \left(1 - 2I\right) \end{pmatrix}$$
For the disease-free equilibrium, the Jacobian has

eigenvalues $\beta$ and $-\alpha$, so the equilibrium is unstable. For the all-infectious equilibrium, the Jacobian has eigenvalues
$ - \beta$  and $ \gamma - \alpha$,
so it is unstable when $\alpha < \gamma$ and stable when $\alpha \geq \gamma$. 
For the interior equilibrium,
the Jacobian 
has eigenvalues
$\frac{-\alpha\beta}{2\gamma} \left(1\pm\sqrt{1+\frac{4\gamma(\alpha-\gamma)}{\beta\alpha}}\right)$.
When $\alpha<\frac{4\gamma^{2}}{\beta+4\gamma}$ the discriminant is negative,
and thus the interior equilibrium is a stable focus. If $\frac{4\gamma^{2}}{\beta+4\gamma} \leq \alpha < \gamma$, the discriminant is real and less than one, 
so the interior equilibrium is a stable node. When $\alpha > \gamma$, the interior equilibrium does not exist, as $I^*_e = \frac{\alpha}{\gamma} > 1$. 
Thus the interior equilibrium is always stable when it exists.
The value of $S$, $I$, and $R$ for the system's unique stable equilibrium for given $\alpha$ is displayed in Fig.  \ref{fig:irSIRSsponbifurcation}.
 
For this model, the initial exponential growth rate is given by $\beta$ (the unstable eigenvalue of the disease-free equilibrium), and then the long-run value of $I(t)$ depends only on the relative values of $\alpha$ and $\gamma$. By contrast, both the exponential growth rate and long-run limit of $I(t)$ depend on $\beta$ and $\gamma$ \cite{keeling2008modeling}, so the irSIRS model is perhaps more capable of separately describing a social phenomenon's catchiness and staying power.

\begin{figure}[h!]
\includegraphics[scale=0.45]{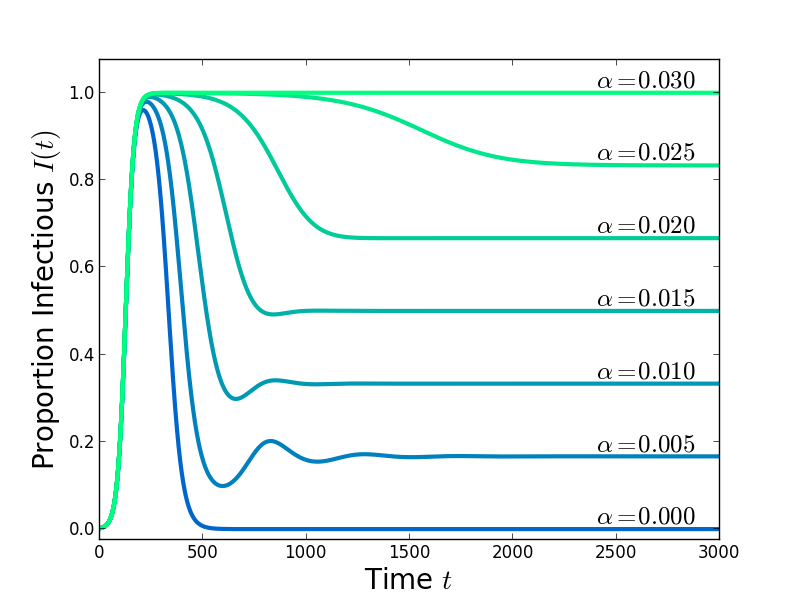}
\caption{irSIRS Model with Spontaneous Loss of Immunity: numerically computed trajectories for $I(t)$ with initial condition $(S(0),I(0),R(0)) = (0.996, 0.002,0.002)$, parameters $\beta = 0.05$ and $\gamma = 0.03$, with $\alpha$ between $0.000$ and $0.030$ with a step size of $0.005$. Trajectories are essentially identical during the exponential growth phase, then diverge near the point of peak infectiousness for the original irSIR model. $I(t)$ converges to 1 when $\alpha = 0.03$, consistent with the analytical result that the all-infectious equlibrium is stable when $\alpha \geq \gamma$. For $\alpha < \frac{4 \gamma^2}{\beta + 4 \gamma}  \approx 0.021$, $I(t)$ oscillates to its long-run equilibrium value.
}
\label{fig:equilibriumvariedirSIRSspon}
\end{figure}

\subsection{Global Stability Analysis}

The analysis of the long time dynamical behavior of a system can be clarified by using a theorem of Bendixson and Dulac, who showed that the existence of certain functions (Dulac functions) imply the absense of periodic orbits. We will now use Dulac functions to show that the locally asymptotic equilibria are globally stable on the interior of the simplex, following the techniques of Hethcote \cite{hethcote1976qualitative}. We reduce Eq. \ref{eq:irSIRSspon} to a system in the variables $I$ and $R$

\begin{equation} \label{eq:irSIRSsponirform}
\begin{array}{ll}
\dsddt{I} &=  \beta( I - I^2) -  \left(\beta + \gamma \right)IR \vspace{2mm}  \\ 
\dsddt{R} &=  \hspace{4mm} \gamma IR - \alpha R  \vspace{2mm} \\ 
\end{array}
\end{equation}
We see that $g(I,R) = \ds\frac{1}{IR}$ is a Dulac function for this system, as $$\dsdel{}{I}\left(g(I,R) \dsddt{I} \right) + \dsdel{}{R}\left(g(I,R) \dsddt{R} \right) = \dsdel{}{I} \left(\frac{\beta}{R} - \frac{\beta I}{R}  - (\beta + \gamma) \right) + \dsdel{}{R}\left( \gamma - \frac{\alpha}{I} \right) = - \frac{\beta}{R} < 0 \: \: \mathrm{when} \: \:  R > 0 $$
so there are no periodic orbits contained in the interior of the simplex, and we apply the Poincare-Bendixson theorem 
to conclude that the unique interior equilibrium (for given $\alpha$) is globally asymptotically stable for initial conditions on the interior of the simplex.

\section{irSIR Model with Demography}

\begin{figure}[h!]
\includegraphics[scale=0.5]{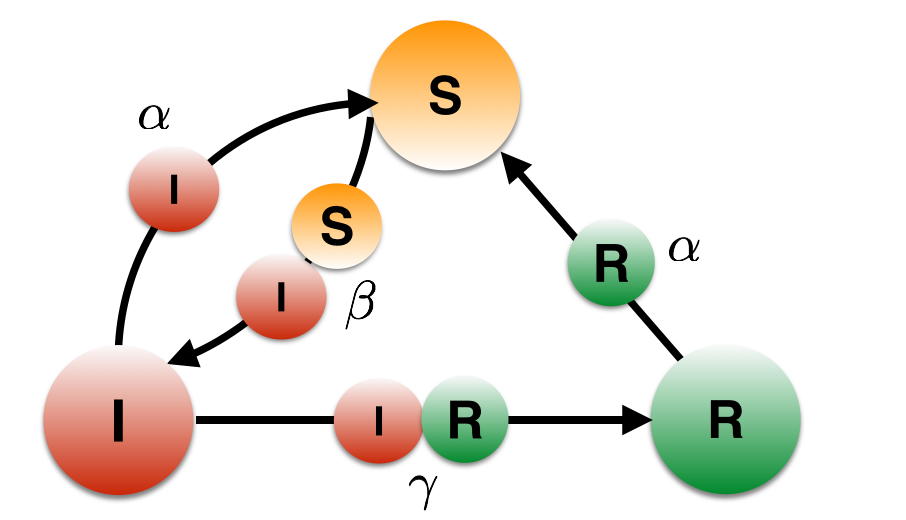}
\caption{Schematic Representation of the irSIR Model with Demography: The $S \to I$ and $I \to R$ transitions are infectious. Individuals in classes $I$ and $R$ are spontaneously replaced by new susceptibles with rate $\alpha$ (i.e. individuals leave the population with rate $\alpha$ and new susceptibles enter the population at the same rate).}
\label{fig:irSIRdemoschematic}
\end{figure}

We extend the irSIR model by considering demographic events that allow new susceptible individuals to be spontaneously introduced to the population and individuals of all types to be spontaneously removed from the population (Fig. \ref{fig:irSIRdemoschematic}). In the epidemiological literature, these events correspond to births and deaths, while Zhu et al motivated the use of demographic terms in OSN models by the creation and deletion of bot accounts \cite{zhu2014demographic}. One additional motivation for adding demographic terms is that individuals can enter and exit the target demographic class for a particular social network. For example, registration for Facebook once required a ``.edu" email address, so the ``birth" and ``death" terms for such a model could be interpreted as the rates of the matriculation of new (and thereby susceptible) students and the graduation of old students. The system of equations is given by

\begin{equation} \label{eq:irSIRdemo}
\begin{array}{ll}
\dsddt{S} &= - \beta SI  + \alpha (1 - S) \vspace{2mm}  \\ 
\dsddt{I} &= \beta SI  - \gamma IR - \alpha I  \vspace{2mm} \\ 
\dsddt{R} &= \gamma IR - \alpha R 
\end{array}
\end{equation}
where parameters satisfy $\beta, \gamma > 0$ and $\alpha \geq 0$. We summarize the analysis of equilibria for this system.

\subsection{Existence of Equilibria}

The system has three equilibria:
\begin{itemize}

\item The disease free equilibrium: $(1,0,0)$.

\item An equilibrium on the $SI$ edge of the simplex: $\left(\frac{\alpha}{\beta}, \frac{\beta - \alpha}{\beta}, 0 \right)$ which only exists when  $ \alpha < \beta$.

\item An interior equilibrium: $\left(\ds\frac{\gamma}{\beta + \gamma}, \ds\frac{\alpha}{\gamma}, \ds\frac{\beta \gamma - \alpha \beta - \alpha \gamma}{\gamma \left(\beta + \gamma \right)} \right)$ which exists for $\alpha < \gamma$ and \begin{equation} \label{eq:eqexists} \beta \gamma - \alpha \beta - \alpha \gamma \geq 0 \Longrightarrow \left(\beta - \alpha \right) \gamma \geq \alpha \beta \end{equation}
It follows from Eq. \ref{eq:eqexists} that $\alpha < \beta$, so the existence of the interior equilibrium on the simplex implies the existence of the edge equilibrium.
To see this, assume $\alpha \geq \beta$. If $\alpha = \beta$, then Eq. \ref{eq:eqexists} implies that $\alpha \beta \leq 0$, which contradicts at least one of $\beta > 0$ or $\alpha \geq 0$. If $\alpha > \beta$,
 we divide both sides of Eq. \ref{eq:eqexists} by the negative quantity $\beta - \alpha$, giving $\gamma \leq \frac{\alpha \beta}{\beta - \alpha} < 0$, contradicting $\gamma > 0$.

\end{itemize}

We can also rewrite the existence condition in Eq. \ref{eq:eqexists} to better represent $\alpha$ as a bifurcation parameter as $$\alpha \leq \frac{\beta \gamma}{\beta + \gamma} = \frac{1}{\frac{1}{\beta} + \frac{1}{\gamma}} \Longleftrightarrow \frac{1}{\alpha} \geq \frac{1}{\beta} + \frac{1}{\gamma}$$

\subsection{Stability Analysis}

We use the simplex constraint to reduce our system to 

\begin{equation} \label{eq:irSIRdemoreduced}
\begin{array}{ll}
\dsddt{S} &= - \beta SI  + \alpha (1 - S) \vspace{2mm}  \\ 
\dsddt{I} &= \left(\beta + \gamma \right) SI - \alpha I - \gamma (I - I^2)  \vspace{2mm} 
\end{array}
\end{equation}
which has Jacobian
$$J(S,I) = \begin{pmatrix} - (\beta I + \alpha) & - \beta S \\ (\beta + \gamma)I & (\beta + \gamma) S - \gamma (1 - 2I) - \alpha \end{pmatrix} $$
For the disease-free equilibrium, the Jacobian has 
eigenvalues $- \alpha$ and $ \beta - \alpha$. so it is asymptotically stable if and only if $\alpha > \beta$. The 
edge equilibrium $(\frac{\alpha}{\beta}, \frac{\beta - \alpha}{\beta}, 0)$ has eigenvalues
$$ \frac{(\beta - \alpha) \gamma - \beta^2}{2} \pm \displaystyle\sqrt{\left(\frac{(\beta - \alpha) \gamma - \beta^2}{2}\right)^2 + (\beta - \alpha) ((\beta - \alpha) \gamma - \alpha \beta)}  $$
For this equilibrium to be asymptotically stable, we need both that
$$ \frac{(\beta - \alpha) \gamma - \beta^2}{2} < 0 
 \Longleftrightarrow \gamma < \frac{\beta^2}{\beta - \alpha} \: \:  \mathrm{and} \: \: 
(\beta - \alpha) ((\beta - \alpha) \gamma - \alpha \beta) < 0 
\Longleftrightarrow \gamma < \frac{\alpha \beta}{\beta - \alpha} < \frac{\beta^2}{\beta - \alpha} $$
where all implications follow from $\alpha < \beta$, and thus the edge equilibrium is stable when $\gamma < \frac{\alpha \beta}{\beta - \alpha}$. The interior equilibrium has eigenvalues
 $ - \frac{\beta \alpha}{2 \gamma} \pm \sqrt{\left( \frac{\beta \alpha}{2 \gamma} \right)^2 + \left( \frac{\beta \alpha}{\gamma} - (\beta - \alpha) \right)} $, 
so it is asymptotically stable if and only if $ \frac{\beta \alpha}{\gamma} - (\beta - \alpha) < 0 
\Longrightarrow \gamma >\frac{\alpha \beta}{\beta - \alpha} $,

where we again applied the condition $\alpha < \beta$. This inequality is also the condition for the existence for the interior equilibrium, so the equilibrium is stable whenever it exists.

\begin{figure}[h!]
\includegraphics[scale=0.45]{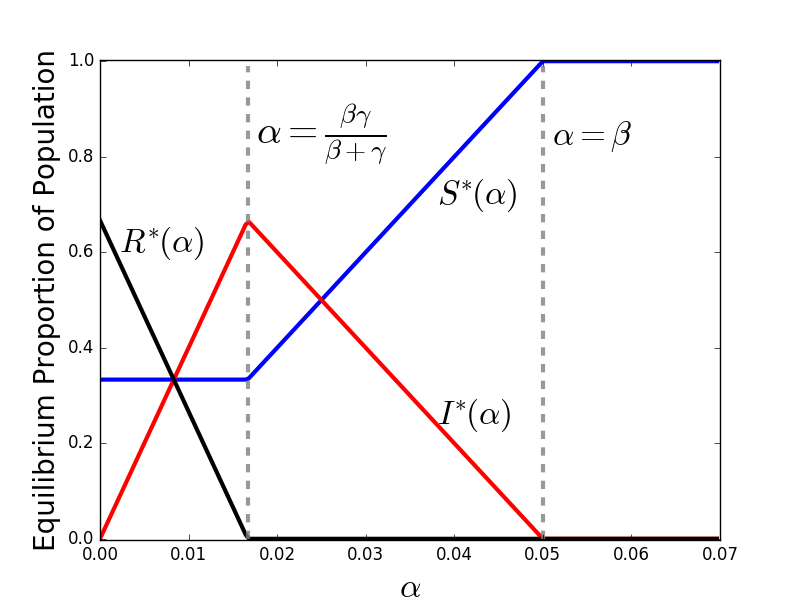}
\caption{Stable Equilibria for irSIR Model with Demography: $S(t)$ (blue), $I(t)$ (red), and $R(t)$ (black) at the unique stable equilibrium point for the given value of $\alpha$. The right gray dotted line corresponds to the bifurcation point where $\alpha = \beta$ and we transition from the edge equilibrium to the disease-free equilibrium. The left gray line corresponds to the point $\alpha = \frac{\beta \gamma}{\beta + \gamma}$, which is the value of $\alpha$ at which we switch from the stable spiral at the interior equilibrium to a stable node at the edge equilibrium. }
\label{fig:irSIRdemobifurcation}
\end{figure}

See  
Fig. \ref{fig:irSIRdemobifurcation} for the unique stable equlibrium for given $\alpha$. $I^*(\alpha)$, the maximum value of $I(t)$ at equilibrium for any $\alpha$, is achieved at the bifurcation point $\alpha = \frac{\beta \gamma}{\beta + \gamma}$. This can also be shown analytically, as we see that $I^*(\alpha)$ increases linearly from $0$ to $\frac{\beta}{\beta + \gamma}$ as $\alpha$ increases from $0$ to $\frac{\beta \gamma}{\beta + \gamma}$ (where the interior equilibrium is stable), and then decreases linearly from $\frac{\beta}{\beta + \gamma}$ to $0$ as $\alpha$ increases from $\frac{\beta \gamma}{\beta + \gamma}$ to $\beta$ (where the edge equilibrium is stable). Therefore $\max_{\alpha}(I^*(\alpha)) = \frac{\beta}{\beta + \gamma}$, and this is achieved for the intermediate demographic rate $\alpha = \frac{\beta}{\beta + \gamma}$. Higher or lower demographic turnover produce smaller proportions of long-run infectious in the social epidemic.

Using the  Dulac function $g(I,R) = (IR)^{-1}$, we can again use the Bendixson-Dulac criterion to rule out periodic orbits on the interior of the simplex, and we conclude that our unique (for given $\alpha$) locally stable fixed points are also globally stable for all initial conditions on the simplex.

\section{Conclusion}

We studied a recently developed class of ODE models for OSN membership that draw inspiration from models in mathematical epidemiology. We used linear stability analysis and peak infection size calculations to characterize the possible behaviors of Cannarella and Spechler's irSIR model. We also considered two extensions of the irSIR model that describe different social mechanisms by which a long-run stable equilibrium with nonzero OSN membership can be achieved by well-mixed populations, and we derived local and global stability results for these extended models. 

In particular, our investigations of the mechanisms of spontaneous loss of immunity and demographic ``births" and ``deaths" have allowed us to show that even a small perturbation of the equations of the irSIR model may result in a long-run globally stable endemic equilibrium, even though the original irSIR model predicts long-run abandonment. Of note, the irSIRS model with spontaneous return to infection allows for an arbitrary fraction of infectious individuals in the long run population and the irSIR model with demography has a peak level of long run infectiousness at an intermediate level of demographic replacement rate $\alpha$.  

The study of the irSIR model and its extensions may also have application to various problems in population dynamics. For example, the coupled terms in the irSIR model bears similarity to generalized Lotka-Volterra model used to describe the ecological dynamics of three species food chains \cite{chauvet2002lotka,rahmani2014analysis,mamat2011numerical}. Exploration of the irSIR model is an interesting exercise because it allows ideas from mathematical epidemiology and nonlinear dynamics to help understand modern human social dynamics, bridging the natural and social sciences and providing inspiration for questions in computer science and the computational social sciences about OSNs and about the spread of social contagion on the internet.

\newpage

\bibliographystyle{unsrt}
\bibliography{SocialNetwork}

\addresseshere

\end{document}